\documentclass[twocolumn,aps,superscriptaddress,raggedbottom,showpacs,floatfix]{revtex4}

\usepackage{amsmath}
\usepackage{amssymb,amsthm}
\usepackage{latexsym}
\usepackage{mathrsfs}
\usepackage{color}
\usepackage{graphicx}
\usepackage{bm}
\usepackage{cancel}

\def\bH{{\bf H}}

\newcommand{\ket}[1]{\ensuremath{\left|{#1}\right\rangle}}

\newcommand{\opa}[1]{\ensuremath{{#1}}}

\newcommand{\tcorder}[1]{\ensuremath{\opa{T_{\cal C}}\left\{{#1}\right\}}}

\newcommand{\copas}[1]{\ensuremath{#1}}

\newcommand{\cds}[1]{\ensuremath{\copas{c}^\dagger_{#1}}}
\newcommand{\ccs}[1]{\ensuremath{\copas{c}_{#1}}}

\newcommand{\nup}{\ensuremath{n_{\uparrow}}}
\newcommand{\ndown}{\ensuremath{n_{\downarrow}}}

\newcommand{\dt}{\ensuremath{\text{d}t}}

\newcommand{\exP}[1]{\ensuremath{\text{exp}\left({#1}\right)}}

\newcommand{\nop}{\ensuremath{\hat{n}}}

\newcommand{\todo}[1]{}
\newcommand{\sqmat}[4]{
\begin{pmatrix}
#1 & #2 \\
#3 & #4
\end{pmatrix}
}

\begin{document}

\title{Multiconfiguration time-dependent Hartree impurity solver for nonequilibrium dynamical mean-field theory}

\author{Karsten Balzer}
\email{karsten.balzer@mpsd.cfel.de}
\affiliation{Max Planck Research Department for Structural Dynamics, University of Hamburg, 22607 Hamburg, Germany}
\affiliation{Center for Free-Electron Laser Science, DESY, Notkestra\ss e 85, 22607 Hamburg, Germany}
\author{Zheng Li}
\affiliation{Center for Free-Electron Laser Science, DESY, Notkestra\ss e 85, 22607 Hamburg, Germany}
\affiliation{Department of Physics, University of Hamburg, 20355 Hamburg, Germany}
\author{Oriol Vendrell}
\email{oriol.vendrell@cfel.de}
\affiliation{Center for Free-Electron Laser Science, DESY, Notkestra\ss e 85, 22607 Hamburg, Germany}
\author{Martin Eckstein}
\affiliation{Max Planck Research Department for Structural Dynamics, University of Hamburg,
22607 Hamburg, Germany}
\affiliation{Center for Free-Electron Laser Science, DESY, Notkestra\ss e 85, 22607 Hamburg, Germany}

%%%%%%%%%%%%%%%%%%%%%%%%%%%%%%%%%%%%%%%%%%%%%%%%%%%%%%%%
% Abstract
%%%%%%%%%%%%%%%%%%%%%%%%%%%%%%%%%%%%%%%%%%%%%%%%%%%%%%%%
\begin{abstract}
Nonequilibrium dynamical mean-field theory (DMFT) solves correlated lattice models by
obtaining their local correlation functions from an effective model consisting of a single
impurity in a self-consistently determined bath. The recently developed mapping of this
impurity problem from the Keldysh time contour onto a time-dependent single-impurity
Anderson model (SIAM) [\textit{C.~Gramsch et al., Phys.~Rev.~B~\textbf{88}, 235106 (2013)}]
allows one to use wave function-based methods in the context of nonequilibrium DMFT.
Within this mapping, long times in the DMFT simulation become accessible by an increasing
number of bath orbitals, which requires efficient representations of the time-dependent
SIAM wave function. These can be achieved by the multiconfiguration time-dependent Hartree
(MCTDH) method and its multi-layer extensions. We find that MCTDH outperforms exact
diagonalization for large baths in which the latter approach is still within reach and
allows for the calculation of SIAMs beyond the system size accessible by exact diagonalization.
Moreover, we illustrate the computation of the self-consistent two-time impurity Green's
function within the MCTDH second quantization representation.
\end{abstract}


%%
%% PACS:
%%
%% 71.27.+a --> Strongly correlated electron systems; heavy fermions
%% 71.10.Fd --> Lattice fermion models (Hubbard model, etc.)
%% 05.70.Ln --> Nonequilibrium and irreversible thermodynamics
%%

\maketitle

%%%%%%%%%%%%%%%%%%%%%%%%%%%%%%%%%%%%%%%%%%%%%%%%%%%%%%%%
% Introduction
%%%%%%%%%%%%%%%%%%%%%%%%%%%%%%%%%%%%%%%%%%%%%%%%%%%%%%%%
\section{Introduction}
Pump-probe experiments with femtosecond time resolution can access the real-time dynamics in materials
with strong correlation effects on the time scale of the 
%ME cavalieri:07 is not really strong correlation. For me, one could omit the reference
electronic motion~\cite{cavalieri:07,wall:11}
and reveal striking phenomena such as photo-induced insulator-to-metal transitions in correlated
Mott and charge-transfer insulators~\cite{perfetti:06,iwai:03} or the pump-induced melting and
recovery of charge density waves~\cite{schmitt:08}. In order to understand the underlying
physical scenario revealed by those experiments, growing theoretical effort has been devoted 
to establish a microscopic description of strongly correlated lattice models out of equilibrium. 
Yet the numerical simulation of quantum many-body systems in nonequilibrium beyond weak-coupling
perturbation theory remains a challenge, in particular for extended systems in dimensions greater 
than one, where the time-dependent density matrix renormalization group (DMRG) method~\cite{schollwoeck:05}
or an exact solution of the Schr\"odinger equation is no longer feasible.

A promising framework to capture both ultrafast dynamics and strong electronic correlation is
the nonequilibrium formulation of dynamical mean-field theory (DMFT)~\cite{schmidt:02,freericks:06,aoki:14},
which generalizes DMFT~\cite{georges:96} to the Keldysh formalism. 
In the framework of DMFT, a lattice model such as the Hubbard model is mapped onto an effective impurity
model, which consists of a single site of the lattice (impurity) coupled to a non-interacting
medium, where electrons are exchanged between the impurity site and the medium. One of the key
developments for advancing DMFT to the nonequilibrium regime is to establish methods to solve
the real-time dynamics of this impurity model far from equilibrium. Impurity solvers that have
been used so far include real-time continuous-time quantum Monte Carlo~\cite{eckstein:09}, which is
numerically exact, but restricted to short times, as well as strong-~\cite{eckstein:10.nca} and
weak-coupling expansions~\cite{tsuji:14}, which have been employed in many studies (see Ref.~\cite{aoki:14} for an overview)
but are restricted to certain parameter regimes. Recently, a Hamiltonian-based impurity solver
scheme has been developed, which further maps the DMFT impurity model onto a 
single-impurity Anderson model (SIAM) with a finite number of bath orbitals~\cite{gramsch:13}.
The latter is then solved to self-consistency by an exact diagonalization method being equivalent 
to time-dependent full configuration interaction (TDCI)~\cite{szaboostlund96},
and one is thus not restricted to either weak or strong on-site Coulomb interaction. 

The mapping of the DMFT impurity model to a SIAM is similar to the related exact-diagonalization
approach to DMFT in equilibrium~\cite{georges:96}, but nevertheless there are important conceptual
differences: Apart from the description of the initial state, the representation of the DMFT bath
can be made {\em exact} for small times, while it requires an increasing number of bath orbitals to
reach longer times~\cite{gramsch:13,balzer:14} (note that other representation schemes might be
useful to obtain qualitatively correct descriptions with few bath orbitals~\cite{hofmann:13} or to
describe the steady state~\cite{arrigoni:13}). Intuitively, increasing the number of bath orbitals
allows the discrete model to develop the finite memory time that is inherent in the original infinite
DMFT bath, i.e., the state can explore a larger Hilbert space without ever returning close to its
initial state. Despite the accuracy of the exact diagonalization method, the exponential scaling of
the Hilbert space dimension as a function of the number of bath orbitals therefore prohibits us to
acquire the dynamics at long time scales. 

Various approaches in different areas of physics have been developed to overcome this course of dimensionality
by finding efficient representations of the wave function. In condensed matter physics, this includes
(time-dependent) DMRG~\cite{schollwoeck:05}, which is based on a matrix product state representation,
and tensor-network representations of many-fermion states~\cite{verstraete:04,vidal:07,murg:07}.
In the present work, we introduce the multiconfiguration time-dependent Hartree (MCTDH) method, which 
has originally been developed for the time propagation of nuclear wave packets in molecular quantum
dynamics~\cite{beck:00,meyer:09}, to treat the real-time dynamics of the SIAM. The MCTDH method provides
a route to represent the wave function with a minimal set of time-dependent basis functions that co-move
with the evolving state. This feature can lead to a tremendous reduction of the configuration space.
Moreover, the more powerful extension of MCTDH, the multi-layer multiconfiguration time-dependent Hartree
(ML-MCTDH) method~\cite{wang:09,manthe:08,vendrell:11}, allows for well-adapted tree-tensor network
decompositions of the many-body wave function.

While MCTDH propagation schemes have recently been used to study transport in the Anderson and 
Anderson-Holstein model~\cite{albrecht:12,wang:13.jchemphys}, the requirements for a nonequilibrium
DMFT calculation are often quite demanding: The Hamiltonian representation of the DMFT impurity model
typically implies strongly time-dependent parameters, the regime of interest includes strong Coulomb
interactions, and, in particular, a fast calculation is required because one needs to perform a large
number of simulations to obtain the impurity Green's function as a function of two time variables. In
order to judge the usefulness of the MCTDH method for nonequilibrium DMFT it is thus important to provide
a comparison of the numerical performance of the method, i.e., to analyze the ability of the ansatz to
compress the SIAM wave function in the typical parameter regime relevant for DMFT and thus to improve
on the exponential increase of the numerical effort on the simulation time which we have described above.
This is one main goal of this paper. 

The article is organized as follows. In Sec.~\ref{sec:theory}, we give a brief overview on nonequilibrium 
DMFT and outline the mapping to a SIAM underlying the Hamiltonian-based impurity solver. We then introduce the 
MCTDH method in Sec.~\ref{sec:mctdh} and discuss its implementation using the Fock space formalism in
Sec.~\ref{sec:fock}. Thereafter, in Sec.~\ref{sec:result}, we present numerical results to assess the performance of
MCTDH as impurity solver in the context of DMFT. Finally, Sec.~\ref{sec:conclusion} provides a general
conclusion.

%%%%%%%%%%%%%%%%%%%%%%%%%%%%%%%%%%%%%%%%%%%%%%%%%%%%%%%%
% Theory
%%%%%%%%%%%%%%%%%%%%%%%%%%%%%%%%%%%%%%%%%%%%%%%%%%%%%%%%
\section{Theoretical Framework}
\label{sec:theory}
In order to combine nonequilibrium DMFT and the multiconfiguration time-dependent Hartree (MCTDH)
method, which we propose as impurity solver, we first give a brief introduction to the DMFT framework.
For a comprehensive introduction to nonequilibrium DMFT and its applications the reader is referred
to Ref.~\cite{aoki:14}.

From a general perspective, we are interested in the real-time evolution of a lattice quantum 
many-body system like the single-band Hubbard model
\begin{align}
\label{eq:ham}
H(t) = \sum_{ij\sigma} t_{ij}(t)\, c_{i\sigma}^\dagger c_{j\sigma}
+
U(t)
\sum_{i}
(n_{i\uparrow}-\tfrac12)
(n_{i\downarrow}-\tfrac12)\,,
\end{align}
which is initially in thermodynamic equilibrium at temperature $T=1/\beta$, and evolves unitarily under
the time-dependent Hamiltonian $H(t)$. In Eq.~(\ref{eq:ham}), the operator $c_{i\sigma}^\dagger$
($c_{i\sigma}$) creates (annihilates) an electron with spin $\sigma$ on site $i$ of the crystal
lattice, $n_{i\sigma}$ is the spin-resolved density, $t_{ij}(t)$ is the hopping matrix element
between sites $i$ and $j$, and $U(t)$ denotes the local Coulomb repulsion. Below, we adopt a
parametrized model where energies (times) are measured in terms of the hopping (inverse hopping)
amplitude ($\hbar=1$). For practical material simulations, $t_{ij}$ and $U$ could in principle be
determined in an {\it ab initio} manner, which is standard for DMFT simulations in equilibrium~\cite{lichtenstein:98,anisimov:97,kotliar:04}.
The unit of time $\hbar/|t_{ij}|$ would be between $20$~fs for narrow-band organic Mott insulators~\cite{wall:11} 
and few fs for transition metal oxides with a bandwidth in the eV range~\cite{imada:98}.

\subsection{Nonequilibrium DMFT and Hamiltonian-based impurity solvers}
\label{sec:dmft}
\begin{figure}
\includegraphics[width=0.2125\textwidth]{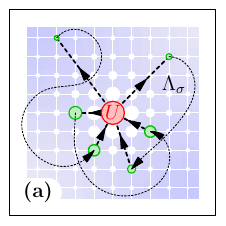}
\includegraphics[width=0.2125\textwidth]{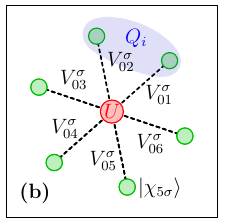}
\caption{
(Color online) \textbf{(a)} DMFT impurity problem according to the action of Eq.~(\ref{eq:dmftaction}).
The arrows illustrate the effect of the two-time hybridization function $\Lambda_\sigma(t,t')$ which
describes all possible processes where a particle of spin $\sigma$ jumps at time $t'$ from the interacting
impurity site (red) to a lattice site $i$ (green), propagates forward or backward in time from site $i$ to
site $j$, and at a time $t$ jumps back to the impurity. \textbf{(b)}~Representation of the DMFT bath by a
single-impurity Anderson model (SIAM) with six bath orbitals and hopping matrix elements $V_{0l}^{\sigma}$,
$l\geq 1$. The variable $Q_1$ illustrates the combination of physical degrees of freedom into combined modes
according to the MCTDH ansatz of Eq.~(\ref{eq2:ansatz2}).}
\label{fig:dmft.mapping}
\end{figure}

The central task of nonequilibrium DMFT based on the Keldysh formalism~\cite{keldysh:64} is to
compute the local contour-ordered Green's function
\begin{align}
 G_\sigma(t,t')=-\mathrm{i}\langle T_{\cal C}\ccs{\sigma}(t)\cds{\sigma}(t')\rangle_{S_\mathrm{loc}}
\end{align}
of an effective single-site impurity model, which exactly replaces the original translationally
invariant lattice problem (\ref{eq:ham}) in the limit of an infinite lattice coordination
(and represents an approximation for finite dimensions). We follow Ref.~\cite{aoki:14} for the 
notation of contour-ordered functions, i.e., time arguments lie on the L-shaped Keldysh
contour ${\cal C}$, and $\langle T_{\cal C} \ldots\rangle_{S_\mathrm{loc}} \equiv \mathrm{Tr} [T_{\cal C} 
e^{S_\mathrm{loc}} \ldots]/\mathrm{Tr} [T_{\cal C} e^{S_\mathrm{loc}}]$ denotes the contour-ordered
expectation value. The action $S_\mathrm{loc}$ of the effective model is illustrated in
Fig.~\ref{fig:dmft.mapping}a and is given by
\begin{multline}
\label{eq:dmftaction}
S_\text{loc}=-\mathrm{i}\int_{\cal C}\mathrm{d}t\biggl[U(t)(\nup(t)-\tfrac12)(\ndown(t)-\tfrac12)-\mu\sum_\sigma n_\sigma(t)\biggr]
\\
-\mathrm{i}\int_{\cal C}
\int_{\cal C}
\mathrm{d}t\,\mathrm{d}t'\sum_{\sigma}\Lambda_{\sigma}(t,t')\cds{\sigma}(t)\ccs{\sigma}(t')\,,
\end{multline}
where the first part contains the Hamiltonian of an isolated site of the original lattice at a
chemical potential $\mu$, and the second part connects the site to a noninteracting continuous
bath which in nonequilibrium is defined by the hybridization function $\Lambda_\sigma(t,t')$. 
In single-site DMFT, the bath must be determined self-consistently from the equations of motion
of $\Lambda_\sigma$ which depend on the impurity Green's function $G_\sigma(t,t')$ and the
time-dependent hopping parameters $t_{ij}(t)$. In the simplest case, for a Bethe lattice in the
limit of an infinite coordination number $Z$ with nearest-neighbor hopping (i.e., semi-elliptical
density of states), the bath is characterized by a self-consistency relation of closed
form~\cite{eckstein:10.epjst},
\begin{equation}
\label{eq:bethe}
\Lambda_\sigma(t,t') =  v(t) G_\sigma(t,t') v(t')\,,
\end{equation}
where the hopping matrix elements in Eq.~(\ref{eq:ham}) are rescaled according to 
$t_{ij}(t)\rightarrow v(t)/\sqrt{Z}$. 

Unfortunately, the DMFT action of the form (\ref{eq:dmftaction}) does not allow for a direct 
solution of the impurity problem with Hamiltonian-based methods. However, an optimal representation
of $S_\mathrm{loc}$ in terms of a time-dependent impurity Hamiltonian with finitely many bath orbitals
can be obtained by a suitable decomposition of the two-time hybridization function. Formally, such
a mapping requires that all impurity correlation functions $\langle{\cal O}(t_1)\ldots\rangle$ 
are the same when computed with the action $S_\mathrm{loc}$ or with the final impurity Hamiltonian $H'(t)$,
i.e.,
\begin{eqnarray}
\label{eq:mapping1}
&&\frac{
\mathrm{Tr}\left(
\tcorder{\exP{S_\text{loc}}  \mathcal{O}(t_1) \ldots} 
\right)
}{
\mathrm{Tr}\left(
\tcorder{\exP{S_\text{loc}}}
\right)
}\nonumber
\\
&\stackrel{!}{=}&
\frac{
\mathrm{Tr}\left(
\tcorder{\exP{-\mathrm{i}\int_{\cal C}\dt H'(t)}  \mathcal{O}(t_1) \ldots} 
\right)
}{
\mathrm{Tr}\left(
\tcorder{\exP{-\mathrm{i}\int_{\cal C}\dt H'(t)}}
\right)
}\,.
\end{eqnarray}
A particularly convenient mapping~\cite{gramsch:13} becomes possible for the single-impurity
Anderson model (SIAM) where $H'=H_\mathrm{SIAM}=H_\mathrm{imp}+H_\mathrm{bath}+H_\mathrm{hyb}$ with
\begin{align}
	\label{eq:ham.siam}
	H_\mathrm{imp}&=-\mu \sum_\sigma n_{0\sigma} + U(t)\left(n_{0\uparrow}-\tfrac12\right)\left(n_{0\downarrow}-\tfrac12\right)\,,\nonumber\\
	H_\mathrm{bath}&=\sum_{l=1}^{L}\sum_{\sigma}(\epsilon_{l\sigma}-\mu) \cds{l\sigma}\ccs{l\sigma}\,,\nonumber\\
	H_\mathrm{hyb}&=\sum_{l=1}^{L}\sum_{\sigma}\left(V^\sigma_{0l}(t)\cds{0\sigma}\ccs{l\sigma}+\text{H.c.}\right)\,.
\end{align}
Here, the impurity site is coupled in a star-pattern by hopping processes of amplitude $V^\sigma_{0l}(t)$
to $L$ individual noninteracting bath orbitals of energy $\epsilon_{l\sigma}$, and the operator $\ccs{l\sigma}$
($\cds{l\sigma}$) annihilates (creates) an electron in a spin-orbital $|\chi_{l\sigma}\rangle$ at bath site $l$
for $l>0$, and at the impurity site for $l=0$ (see illustration in Fig.~\ref{fig:dmft.mapping}b).
The hybridization function of the SIAM is given by
\begin{align}
\label{eq:siam.hybfct}
 \Lambda_\sigma'(t,t')=\sum_{l=1}^{L}V_{0l}^\sigma(t)g(\epsilon_{l\sigma},t,t')V_{l0}^\sigma(t')\,,
\end{align}
where $g(\epsilon,t,t')=-\mathrm{i}[\theta_{\cal C}(t,t')-f(\epsilon)]\mathrm{e}^{-\mathrm{i}\epsilon(t-t')}$
is the Green's function of an isolated bath orbital, $f(\epsilon)=1/(\mathrm{e}^{\beta\epsilon}+1)$ denotes 
the Fermi distribution, and $\theta_{\cal C}$ is the contour step function. Since the exponential 
$\mathrm{e}^{-\mathrm{i}\epsilon(t-t')}$ can be absorbed into the time-dependence of the parameters
$V_{0l}^\sigma(t)$, the problem of representing the DMFT action by the Hamiltonian~\eqref{eq:ham.siam}
has thus been reduced to a factorization of a ``two-time matrix'' $\Lambda_\sigma(t,t')$ in terms of
time-dependent functions.

If the bath is initially decoupled from the impurity (this is commonly referred to as the atomic
limit), the initial state of the system is entirely described in terms of the impurity density matrix,
and $\Lambda_\sigma(t,t')$ is only nonzero for times $t,t'>0$ on the real part of the contour. The 
parameters in the SIAM can then be obtained by demanding that the greater and lesser components of 
the original hybridization function $\Lambda_\sigma(t,t')$ and $\Lambda_\sigma'(t,t')$ of Eq.~(\ref{eq:siam.hybfct}) 
are identical for all times $t$ and $t'$. In practice, this leads to a matrix decomposition of 
$\Lambda_\sigma(t,t')$, where the matrix rank $N_\mathrm{t}$ is defined by the discretization of
the times $t$ and $t'$ according to $0,\delta t, 2\delta t,\ldots,(N_\mathrm{t}-1)\delta t$. By
choosing the bath energies of the SIAM such that the occupations $f(\epsilon_{l\sigma}-\mu)$ are
either $0$ or $1$, the greater and lesser components can then be decomposed independently of one another:
\begin{align}
\label{eq:lambda.decomp}
 -\mathrm{i}\Lambda^<_\sigma(t,t')&=\sum_{l=1}^{L/2}V_{0l}^\sigma(t)[V_{0l}^\sigma(t')]^*\,,&\\
 \mathrm{i}\Lambda^>_\sigma(t,t')&=\sum_{l=L/2+1}^{L}V_{0l}^\sigma(t)[V_{0l}^\sigma(t')]^*\,,\nonumber
\end{align}
where we have initially occupied the first half of the spin-orbitals $|\chi_{l\sigma}\rangle$ and left the
other half empty (note that non-uniform partitions are also possible).

It is obvious that Eq.~(\ref{eq:lambda.decomp}) holds in general only in the limit $L\rightarrow\infty$,
ensuing from an infinite DMFT bath. However, appropriate representations can usually be obtained already
for a rather small number of bath orbitals~\cite{balzer:14}. Furthermore, by using a low-rank Cholesky
approximation in Eq.~(\ref{eq:lambda.decomp}) one can guarantee that the representation of the hybridization
function is always correct at short times such that a gradual increase of $L$ allows us to successively
approach longer and longer simulation times \cite{gramsch:13}. Finally, we note that if impurity and bath
hybridize already in the initial state (i.e., for $t\leq0$ where the Hamiltonian $H(t)$ is time-independent),
the SIAM representation of the DMFT action (\ref{eq:dmftaction}) requires additional bath sites which
describe the time evolution of initial correlations, for details see also~\cite{gramsch:13}.

\subsection{Multiconfiguration time-dependent Hartree}
\label{sec:mctdh}
Practical applications of DMFT require an efficient solver for the time-dependent
Schr{\"o}dinger equation (TDSE) of the impurity model~(\ref{eq:ham.siam}).
The MCTDH method~\cite{meyer:90,beck:00,meyer:09,mctdh:package}, which has been
applied to a variety of molecular quantum dynamics problems since its inception
more than $20$ years ago, is a general approach to efficiently solve the TDSE for
multidimensional systems that tries to alleviate the exponential increase of
computational effort with system size by a compact representation of the time-dependent 
state vector of the system. We will first describe the original formulation
of MCTDH for systems of distinguishable degrees of freedom. The extension to 
indistinguishable particles (Sec.~\ref{sec:fock}) is then very similar and even
uses essentially the same numerical implementation. 

The standard wave function ansatz to solve the TDSE for a system with $f$ degrees
of freedom reads
\begin{eqnarray}
\label{eq2:ansatz0}
  \nonumber
  |\Psi(q_1,\dots,q_f,t)\rangle & = &
  \sum_{j_1}^{N_1}\ldots\sum_{j_f}^{N_f}{C_{j_1\ldots j_f}(t)
     \prod_{\kappa=1}^{f}{|\chi_{j_{\kappa}}^{(\kappa)}(q_{\kappa})\rangle}} \\
  & = &
     \sum_J C_J(t)\, |\Xi_J\rangle\,,
  \end{eqnarray}
which expands the wave function as a sum of Hartree products of one-dimensional
primitive basis functions $|\chi_{j_{\kappa}}^{(\kappa)}(q_{\kappa})\rangle$. For
convenience, these functions are chosen orthonormal without loss of generality.
Once a primitive basis has been selected, the time evolution of the system is fully
determined by the set of time-dependent expansion coefficients $C_{j_1\ldots j_f}(t)$,
which constitute a multidimensional tensor of rank $f$. Inserting Eq.~(\ref{eq2:ansatz0})
into the TDSE and multiplying from the left by $\langle\Xi_L|$ results in the linear
matrix equation
\begin{equation}
  \label{eq:eom0}
  \mathrm{i}\dot{C}_L = \sum_{J}\langle\Xi_L\left|H\right|\Xi_J\rangle C_J\,.
\end{equation}
The standard approach has an exponential scaling $N^f$ in the number of expansion
coefficients with the dimensionality and is therefore only practicable for a few degrees of freedom.

The MCTDH ansatz for the wave function reads
\begin{eqnarray}
\label{eq2:ansatz1}
    \nonumber
    |\Psi(q_1,\dots,q_f,t)\rangle & = &
    \sum_{j_1}^{n_1}\ldots\sum_{j_f}^{n_f}{A_{j_1\ldots j_f}(t)
    \prod_{\kappa=1}^{f}{|\varphi_{j_{\kappa}}^{(\kappa)}(q_{\kappa},t)\rangle}} \\
    & = &  \sum_{J}{A_J(t)|\Phi_J(t)\rangle}\,,
\end{eqnarray}
where the key difference to the standard ansatz is the introduction of time-dependent
single-particle functions (SPF) $|\varphi_{j_{\kappa}}^{(\kappa)}(q_{\kappa},t)\rangle$,
which are taken to be orthonormal for all times. The MCTDH equations of motion for the
time-dependent coefficients and SPFs are derived from the Dirac-Frenkel variation principle
and read~\cite{beck:00}
\begin{eqnarray}
\label{eq:eom}
\mathrm{i}\dot{A}_J &=& \sum_{L}\langle\Phi_J\left|H\right|\Phi_L\rangle A_L\,,\\
\mathrm{i}\dot{{\bf \varphi}}^{(\kappa)}
      &=& (1-P^{(\kappa)})({\bf\rho}^{(\kappa)})^{-1}\langle\bH\rangle^{(\kappa)}
      {\bf \varphi}^{(\kappa)}\,.\nonumber
\end{eqnarray}
Here a vector notation
${\bf  \varphi}^{(\kappa)}=(|\varphi_1^{(\kappa)}\rangle,\ldots,|\varphi_{n_{\kappa}}^{(\kappa)}\rangle)^T$
is used,
\begin{equation}
  \label{eq:projector}
  P^{(\kappa)}=
 \sum_{j=1}^{n_{\kappa}}|{\bf \varphi}^{(\kappa)}_j\rangle\langle{\bf \varphi}^{(\kappa)}_j|\,
\end{equation}
is the projector on the space spanned by the SPFs for the $\kappa$th degree of
freedom, and $\langle \bH\rangle^{(\kappa)}$ and $\rho^{(\kappa)}$ are mean-fields
and the density matrix. By defining single-hole functions $|\Psi_l^{(\kappa)}\rangle$
as linear combinations of Hartree products of $(f-1)$ SPFs without the SPFs for
the $\kappa$th degree of freedom $q_{\kappa}$,
\begin{eqnarray}
\label{eq:single_hole_func}
|\Psi_l^{(\kappa)}\rangle&=&\sum_{j_1}\ldots\sum_{j_{\kappa-1}}\sum_{j_{\kappa+1}}\ldots\sum_{j_f}
A_{j_1\ldots j_{\kappa-1}lj_{\kappa+1}\ldots j_f}\nonumber\\
&&\times\,|\varphi_{j_1}^{(1)}\rangle\ldots|\varphi_{j_{\kappa-1}}^{(\kappa-1)}\rangle|\varphi_{j_{\kappa+1}}^{(\kappa+1)}\rangle\ldots|\varphi_{j_f}^{(f)}\rangle\,,
\end{eqnarray}
one can write $\langle \bH\rangle^{(\kappa)}$ and $\rho^{(\kappa)}$
in compact forms as
\begin{equation}
\label{eq:mean_fields}
\langle H\rangle^{(\kappa)}_{jl}=\langle\Psi_j^{(\kappa)}| H |\Psi_l^{(\kappa)}\rangle
\end{equation}
and
\begin{eqnarray}
\rho^{(\kappa)}_{jl}&=
%%
%% ME: check!
%%
&\langle\Psi_j^{(\kappa)}|\Psi_l^{(\kappa)}\rangle
%&|\Psi_j^{(\kappa)}\rangle\langle\Psi_l^{(\kappa)}|
=\sum_{j_1}\ldots\sum_{j_{\kappa-1}}\sum_{j_{\kappa+1}}\ldots\sum_{j_f}\nonumber\\
&&A^*_{j_1\ldots j_{\kappa-1}jj_{\kappa+1}\ldots j_f} A_{j_1\ldots j_{\kappa-1}lj_{\kappa+1}\ldots j_f}\,.
\end{eqnarray}

For a complete set of SPFs, $n_\kappa = N_\kappa$, one has $P^{(\kappa)}=1$ and hence 
$\mathrm{i}\dot{{\bf \varphi}}^{(\kappa)}=0$, such that Eqs.~\eqref{eq:eom} correspond
to the standard method of Eq.~\eqref{eq:eom0}. With variationally optimal SPFs, the number
of basis functions per degree of freedom can be kept smaller than the number of
time-independent primitive functions. The number of coefficients in the $A$-vector still
grows exponentially as $n^f$, but now to a smaller base (assuming equal $N_\kappa$ and 
$n_\kappa$ for all degrees of freedom, the number of time-dependent coefficients representing
the MCTDH wave function is $n^f + f N n$). In the limit of all $n_\kappa=1$, the evolution
is described by a single Hartree product of time-evolving SPFs, which corresponds to the
time-dependent Hartree method. The advantage of MCTDH over the standard method lies on the
fact that a much smaller number of differential equations has to be solved and the accuracy
and cost of the calculation can be controlled by choosing the number of SPFs $n_\kappa$ for
each degree of freedom. However, both equations in~(\ref{eq:eom}) are non-linear, and
$\langle\Phi_J\left|H\right|\Phi_L\rangle$ as well as the mean fields $\langle\bH\rangle^{(\kappa)}$
must be rebuilt at every time step, which is usually the largest computational burden of
MCTDH calculations.

\begin{figure}
 \includegraphics[width=0.4\textwidth]{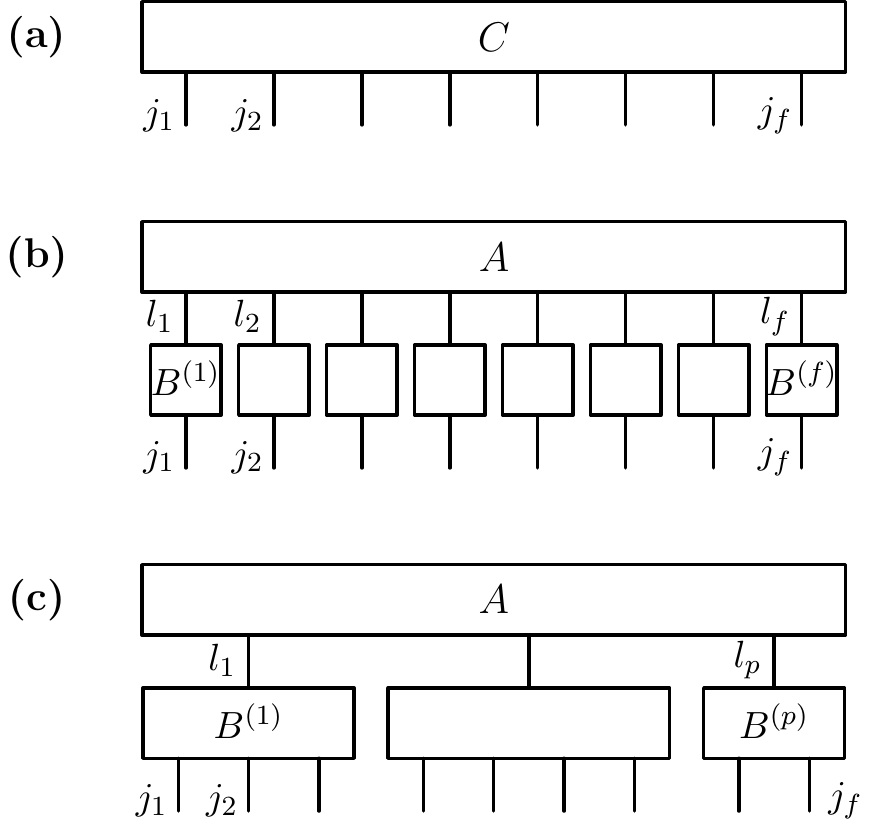}
\caption{
From top to bottom, tensor networks representing the coefficients of the
MCTDH ans{\"a}tze of Eqs.~(\ref{eq2:ansatz0}), (\ref{eq:tensor1}) and (\ref{eq:tensor2}),
respectively. Each box represents a tensor with as many indices as outgoing lines. A sum
is performed over all indices at a line connecting two tensors.}
\label{fig:tensor}
\end{figure}

Although other choices may be possible, the SPFs are often expanded on a set of
time-independent orthogonal basis functions as in the standard approach, i.e.,
\begin{equation}
    \label{eq2:spf}
    |\varphi_{j_\kappa}^{(\kappa)}(q_\kappa,t)\rangle =
    \sum_{l_\kappa}^{N_\kappa}
    B^{(\kappa)}_{j_\kappa l_\kappa} |\chi_{l_\kappa}^{(\kappa)}(q_\kappa)\rangle\,.
\end{equation}
The MCTDH ansatz can then be regarded as a way to compactify the $C_{j_1\ldots j_f}$
tensor introduced in Eq.~(\ref{eq2:ansatz0}) as
\begin{equation}
    \label{eq:tensor1}
    C_{j_1 \ldots j_f} = \sum_{l_1}^{n_1}\ldots\sum_{l_f}^{n_f}
    A_{l_1 \ldots l_f} \prod_{\kappa=1}^{f} B^{(\kappa)}_{l_\kappa j_\kappa}\,.
\end{equation}
This decomposition, which is graphically represented in Fig.~\ref{fig:tensor}b,
is known as the Tucker tensor decomposition~\cite{tucker:66} and has the same form
as a matrix singular value decomposition generalized to the multidimensional
case~\cite{kolda:09}. The dimensionality of the $A$-vector can be further 
reduced by combining physical coordinates $q$ in fewer combined modes or logical
coordinates $Q$. In this case the ansatz reads
\begin{eqnarray}
\label{eq2:ansatz2} |\Psi(q_1,\dots,q_f,t)\rangle &\equiv& |\Psi(Q_1,\dots,Q_p,t)\rangle \nonumber\\
                      &  =  & \sum_{j_1}^{n_1}\ldots\sum_{j_p}^{n_p}{A_{j_1\ldots j_p}(t)\prod_{\kappa=1}^{p}{|\varphi_{j_{\kappa}}^{(\kappa)}(Q_{\kappa},t)}\rangle} \nonumber \\
                      & = & \sum_{J}{A_J(t)|\Phi_J(t)\rangle}\,,
\end{eqnarray}
where $p$ is the number of MCTDH combined modes and the SPFs are now multidimensional
functions. In terms of a more general tensorial decomposition similar to Eq.~(\ref{eq:tensor1}),
the mode combination ansatz becomes (cf.~Fig.~\ref{fig:tensor}c)
\begin{equation}
    \label{eq:tensor2}
    C_{j_1\ldots j_f} = \sum_{l_1}^{n_1}\ldots\sum_{l_p}^{n_p}
    A_{l_1\ldots l_p} \prod_{\kappa=1}^{p} B^{(\kappa)}_{l_\kappa; J_\kappa}\,,
\end{equation}
where $J_\kappa$ refers to a multi-index collecting the  $d_\kappa$ degrees of freedom that
are combined into one combined mode $\kappa$, i.e.,
$J_1=(j_1,\ldots,j_{d_1})$,
$J_2=(j_{d_1+1},\ldots,j_{d_1+d_2})$, $\ldots$,
$J_p=(j_{d_1+\ldots+d_{p-1}+1},\ldots,j_{f})$.
The use of mode combination allows one to distribute the cost of the calculation in an
optimal way between the propagation of the $A$ and $B$ coefficients (i.e., between the
$A$-vector and the SPFs). Large combined modes lead to a small $A$-vector that can be
efficiently propagated but result in the costly propagation of multidimensional SPFs.
Providing a scheme for efficiently propagating multidimensional wave functions is what MCTDH
does in the first place by introducing a multiconfigurational ansatz. A natural
extension consists in expanding the multidimensional SPFs as sums of products of
time-dependent basis functions of lower dimensionality, what is known as multi-layer MCTDH
(ML-MCTDH)~\cite{wang:09,manthe:08,vendrell:11}. In terms of the tensorial argument, it 
consists in decomposing the $B^{(\kappa)}_{l_\kappa;J_\kappa}$-tensors in the same form 
as Eq.~(\ref{eq:tensor1}) or Eq.~(\ref{eq:tensor2}).

Computationally, MCTDH is most efficient when the system Hamiltonian is given by a sum of 
products of low-dimensional operators, as this immediately factorizes the matrix elements in
Eqs.~(\ref{eq:eom}) into products of lower-dimensional integrals (note that the Hamiltonian and
mean-field matrix elements need to be reevaluated at every time step because they depend
on the wave function via the time-dependent SPFs). For model Hamiltonians fulfilling this 
product form (ML-)MCTDH has been applied to thousands of degrees of 
freedom, e.g.,~\cite{wang:08.njp}.

\subsection{MCTDH in second quantization representation}
\label{sec:fock}
Using Hartree products as the elementary configurations, the MCTDH framework introduced
above rather describes the time evolution of {\it distinguishable particles} than the
dynamics of fermionic or bosonic many-body states. Approaches which explicitly account for
the exchange symmetry of the wave function are MCTDH for fermions (MCTDHF)~\cite{caillat:05,nest:09,ulusoy:12,zhang:14,hochstuhl:14},
which is based on a multiconfiguration expansion of the wave function in terms of Slater 
determinants built from time-dependent spin-orbitals, and the bosonic version (MCTDHB), in 
which the many-body configurations are taken to be permanents~\cite{alon:07,alon:08,cao:13}.

It is possible, however, to describe the dynamics of many-body systems of fermionic
or bosonic symmetry by working explicitly within the occupation number representation,
such that the state vectors are members of Fock space instead of a Hilbert space. 
This scheme was introduced by Thoss and Wang under the name MCTDH in second quantization
representation (MCTDH-SQR)~\cite{wang:09}.  As we will use MCTDH-SQR instead of MCTDHF to
solve the DMFT impurity problem with $N$ fermionic particles, we shortly review the key
points of the scheme and refer the interested reader to the original reference.

For fermions and $M$ spin-orbitals, the basis of the Fock space is given by
\begin{equation}
\ket{n_1,n_2,\ldots,n_M}=\prod_{P=1}^{M}(c_P^\dagger)^{n_P}\ket{0_1,0_2,\ldots,0_M}\,,
\end{equation}
where $n_P=0,1$ are the allowed occupations of the spin-orbital $\ket{\chi_{P}}$, $\ket{0_1,0_2,\ldots,0_M}$ 
denotes the empty vacuum state, and $\cds{P}$ denotes the fermionic creation operator, satisfying
anticommutation relations with the associated annihilation operator
$\ccs{P}$,
\begin{eqnarray}
\{\ccs{P},\cds{Q}\}&\equiv&\ccs{P}\cds{Q}+\cds{Q}\ccs{P}=\delta_{PQ}\,,\nonumber\\
\{\ccs{P},\ccs{Q}\}&=&\{\cds{P},\cds{Q}\}=0\,. 
\end{eqnarray}
The key step is a Jordan-Wigner transformation of the fermionic degrees of freedom, i.e.,
the basis vectors in the occupation number representation are represented as a Hartree
product of kets $|n_P\rangle$ for the occupation of each spin-orbital,
\begin{equation}
\label{eq:fock_hartree}
\ket{n_1,n_2,\dots,n_M} \equiv
|n_1\rangle \otimes |n_2\rangle \otimes \cdots \otimes |n_M\rangle\,.
\end{equation}
Regarding Eq.~(\ref{eq2:ansatz0}), the expansion coefficients $C_{n_1\ldots n_M}$ then do not fulfill any particular
antisymmetry relation upon exchange of their indices and thus can be compactified
according to Eq.~(\ref{eq:tensor2}), leading to the standard MCTDH for distinguishable
degrees of freedom.

For actual manipulations, the occupation number states of a spin-orbital $\ket{\chi_j}$ are
mapped onto a two-dimensional vector space
\begin{eqnarray}
\label{eq:fock_basis}
\ket{n_j=0}&\Leftrightarrow& \binom{0}{1}\,,\nonumber\\
\ket{n_j=1}&\Leftrightarrow& \binom{1}{0}\,,
\end{eqnarray}
while creation and annihilation and all derived operators acting in Fock space
are mapped to products of $2\times 2$ matrices, i.e.,
\begin{eqnarray}
    \hat{c}^\dagger_{P}&\equiv&\left(\prod_{Q=1}^{P-1}(-1)^{n_Q}\right)
    \tilde{\hat{c}}^\dagger_{P}
    =\left(\prod_{Q=1}^{P-1}\hat{S}_Q\right)\tilde{\hat{c}}^\dagger_{P}\,,\nonumber\\
    \tilde{\hat{c}}^\dagger_{P}&\equiv&\sqmat{0}{1}{0}{0}\,,\nonumber\\
\hat{S}_P&=&(-1)^{n_P}\equiv\sqmat{1}{0}{0}{-1}\,. 
\end{eqnarray}
The matrix form of the annihilation operator results from the Hermitian conjugate
of $c^\dagger_{P}$, 
\begin{eqnarray}
    \hat{c}_{P}&\equiv&\left(\prod_{Q=1}^{P-1}(-1)^{n_Q}\right)\tilde{\hat{c}}_{P}\,,\nonumber\\
    \tilde{\hat{c}}_{P}&\equiv&\sqmat{0}{0}{1}{0}\,. 
\end{eqnarray}
Moreover, the number of electrons in a Fock state $\ket{n_1,n_2,\dots,n_M}$ is given by
$N=\sum_{P=1}^M n_P$, and the electron number operator $\nop_P$ for spin
orbital $\ket{\chi_{P}}$ can be written as 
\begin{equation}
 \hat{n}_P\equiv\sqmat{1}{0}{0}{0}\;.
\end{equation}

It is clear that any operator in second quantization, such as the Hamiltonians in
Eqs.~(\ref{eq:ham}) and (\ref{eq:ham.siam}), consists of a sum of products of terms acting
on one degree of freedom only, where the degrees of freedom are the occupation numbers of
each spin-orbital. All the exchange symmetry logic is contained in the products of sign-change
operators $S_Q$ acting on the degrees of freedom in front of position $P$ where a particle is being
either created or annihilated, i.e., the (anti)symmetry properties of the system are
carried by the operator and not by the state vector as is the case in first quantization.

In practice, virtually any MCTDH implementation for distinguishable particles with the
possibility of mode combination or multi-layer MCTDH, e.g., the Heidelberg MCTDH package
used here~\cite{mctdh:package}, can perform MCTDH-SQR calculations without further
modification. All that needs to be done is to define the corresponding system Hamiltonian
making use of the representation of second quantization operators as
(products of) $2\times 2$ matrices according to the rules introduced above.

%%%%%%%%%%%%%%%%%%%%%%%%%%%%%%%%%%%%%%%%%%%%%%%%%%%%%%%
% Results and discussion
%%%%%%%%%%%%%%%%%%%%%%%%%%%%%%%%%%%%%%%%%%%%%%%%%%%%%%%
\section{Results and discussion}
\label{sec:result}
In this section, we apply the MCTDH-SQR method as impurity solver and evaluate its performance on
the basis of the computational cost. In the first part, we outline the procedure for a simple
test bath and compute the time-dependent wave function of the corresponding single-impurity Anderson
model (SIAM) including $L$ bath sites and $N=N_\uparrow+N_\downarrow=L+1$ fermions for various on-site
interactions $U$. In the second part, we discuss the self-consistency and illustrate the computation
of the two-time impurity Green's function.

\begin{figure}[tbp]
 \includegraphics[width=0.485\textwidth]{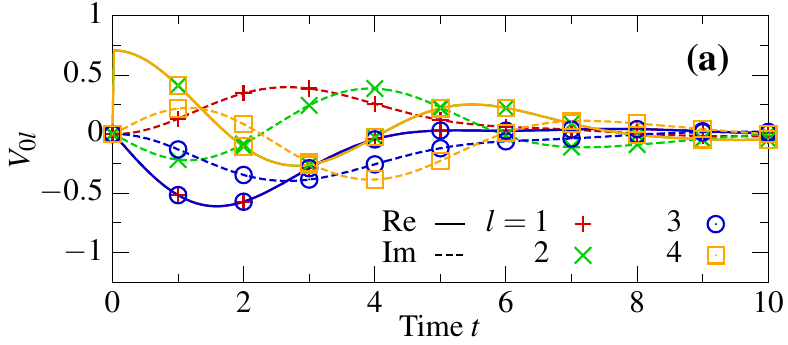}
 \includegraphics[width=0.485\textwidth]{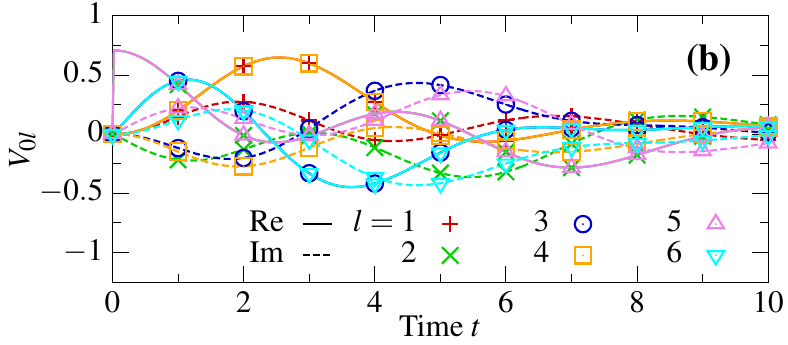}
 \caption{(Color online) Time evolution of the complex hopping matrix elements $V_{0l}(t)=V_{0l}^\uparrow(t)=V_{0l}^\downarrow(t)$
 for a SIAM with \textbf{(a)} $L=4$ and \textbf{(b)} $L=6$ bath orbitals and a reference bath which is
 governed by the equilibrium Green's function of Eq.~(\ref{eq:bathgf}) with inverse temperature $\beta=1$.}
 \label{fig:siam.hopping.parameters}
\end{figure}

\subsection{Model setup}
\label{subsec:typ.siam-model}
To assess the performance of MCTDH-SQR for a time-dependent impurity problem which is representative
for a DMFT calculation, we solve an impurity model which is suddenly coupled to a  bath with semi-elliptical density 
of states and temperature $T=\beta^{-1}=1$, i.e., we choose a hybridization function $\Lambda_\sigma(t,t')=v(t)g_\sigma(t,t')v(t')$,
where the coupling $v(t)$ to the bath is given by a Heavyside step function, and $g_\sigma$ is the 
equilibrium Green's function of the uncoupled bath,
\begin{align}
\label{eq:bathgf}
g_\sigma^\gtrless(t,t')=\mp\mathrm{i}\int d\omega f^\gtrless(\omega)A(\omega)\mathrm{e}^{-\mathrm{i}\omega(t-t')}\,, 
\end{align}
with $f^<(\omega)=f(\omega)=1/(\mathrm{e}^{\beta\omega}+1)$, $f^>(\omega)=1-f(\omega)$, and 
$A(\omega)=\tfrac{1}{2\pi}\sqrt{4-\omega^2}$. 
The complex hopping parameters $V_{0l}^\sigma(t)$ in the SIAM then follow from a low-rank
Cholesky decomposition of $\Lambda_\sigma(t,t')$~\cite{gramsch:13}. Figure~\ref{fig:siam.hopping.parameters} shows
the resulting hopping parameters for a setup with $L=4$ and $L=6$ bath sites on a time window up
to $t=10$; the time discretization comprises $n_t=500$ time steps.

For the MCTDH-SQR setup, we group two bath sites (i.e., four bath spin-orbitals
corresponding to four degrees of freedom) into one combined mode. As each spin-orbital
can be in either state $|0\rangle$ (empty) or $|1\rangle$ (occupied), the span with
$2^4=16$ SPFs represents the full Fock space. The impurity is left as a separate
mode, which will in practice always be described with the maximum
of $2^2=4$ SPFs. Starting from the atomic limit, the impurity site is initially
decoupled from the bath and is occupied by a single up- or down-spin electron.
Consistent with the decomposition scheme outlined in Sec.~\ref{sec:dmft}, the
bath orbitals have different initial populations: The first half is doubly occupied
whereas the second half is empty ($L$ even). 

\begin{figure}
 \includegraphics[width=0.485\textwidth]{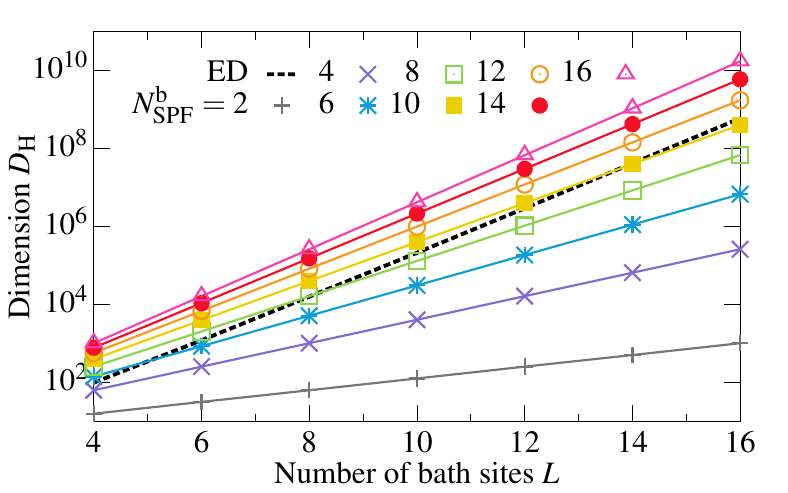}
 \caption{(Color online) Hilbert space dimension of the SIAM as function of the number of bath sites
 $L$ (black dashed line) and dimensionality of the corresponding $A$-vectors in MCTDH-SQR (colored lines)
 for a setup where four bath spin-orbitals are treated in a combined mode ($N_\mathrm{CM}^\mathrm{b}$=4)
 and $N_\mathrm{SPF}^\mathrm{b}$ single-particle functions are involved, cf.~Eq.~(\ref{eq:dim.mctdh}).}
 \label{fig:dimension.scaling}
\end{figure}

Assuming a SIAM with $N_\uparrow=L/2+1$ spin-up particles and $N_\downarrow=L/2$ 
spin-down particles, the dimension of the  Hilbert space of the SIAM is given by 
\begin{align}
\label{eq:dim.ed}
 D_\mathrm{H}=\binom{L+1}{L/2+1}\binom{L+1}{L/2}\,.
\end{align}
On the other hand, the $A$-vector $A_{j_1\ldots j_p}$ in the MCTDH ansatz of
Eq.~(\ref{eq2:ansatz2}) has dimension
\begin{align}
\label{eq:dim.mctdh}
 D_\mathrm{A}=2^2(N_\mathrm{SPF}^\mathrm{b})^{2L/N_\mathrm{CM}^\mathrm{b}}\,,
\end{align}
for an orbital partition scheme with $N_\mathrm{CM}^\mathrm{b}$ bath spin-orbitals in a 
combined mode and each combined mode being represented by $N_\mathrm{SPF}^\mathrm{b}$ SPFs 
(as stated above the impurity degrees of freedom are treated in a single separate mode and are accounted 
for by the factor $2^2$). From Fig.~\ref{fig:dimension.scaling} we observe that (despite the 
exponential scaling of the configuration space with $L$) the application of MCTDH-SQR 
can become favorable against exact diagonalization (ED) for specific numbers of SPFs 
$N_\mathrm{SPF}^\mathrm{b}$ at fixed $L$, provided that the relevant observables of the 
impurity model are satisfactorily resolved in time. Note also that for large numbers of
SPFs the size of the $A$-vector can exceed the size of the Hilbert space of the SIAM. This
is due to the fact that MCTDH-SQR is defined in the whole Fock space and an exact
calculation corresponds to the sum of all Hilbert space sizes corresponding to all
possible occupations.

\begin{figure}
 \includegraphics[width=0.485\textwidth]{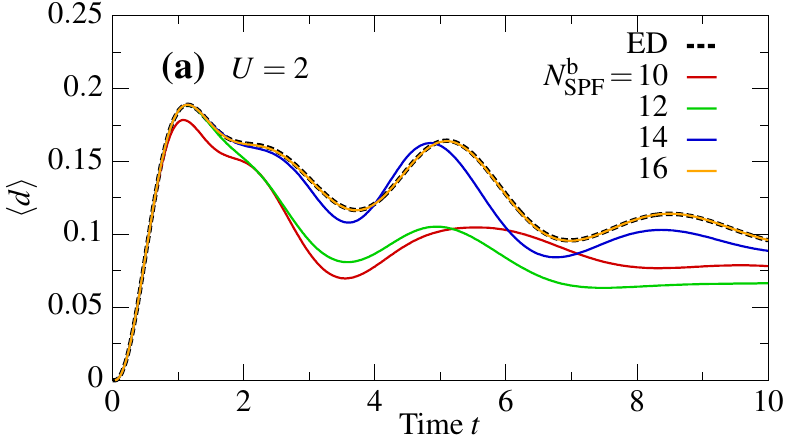}
 \includegraphics[width=0.485\textwidth]{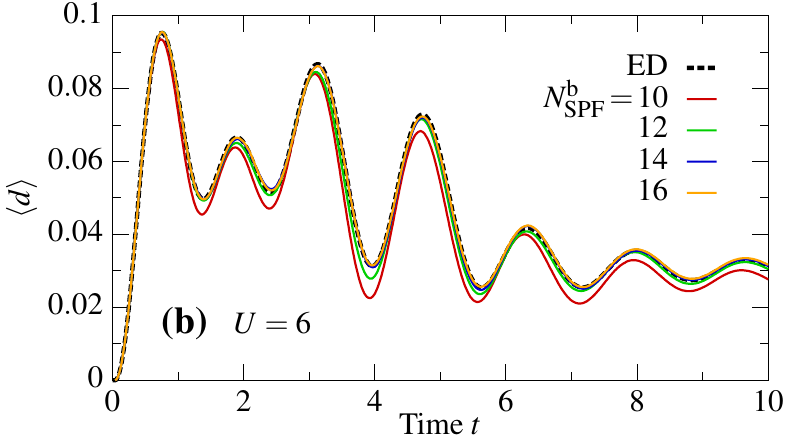}
 \caption{(Color online) Time-dependent double occupancy $\langle d\rangle(t)$ of the impurity site for
 the SIAM with $L=4$ bath orbitals at \textbf{(a)}~$U=2$ and \textbf{(b)}~$U=6$, calculated by exact
 diagonalization (ED) and MCTDH-SQR with various numbers of SPFs ($N_\mathrm{SPF}^\mathrm{b}$).}
 \label{fig:siam.double.occupation}
\end{figure}

\subsection{Comparison to exact diagonalization}
\label{subsec:typ.siam}

\subsubsection{Time evolution of the double occupancy}
To examine the quality of the MCTDH ansatz of the SIAM for different numbers of SPFs,
we compute the time-dependent impurity double occupancy
\begin{align}
\langle d\rangle(t)=\langle \Psi(t)|n_{0\uparrow}n_{0\downarrow}|\Psi(t)\rangle 
\end{align}
for various sizes $L$ of the bath and different on-site interactions and compare it to exact
reference data which is obtained by ED. In Fig.~\ref{fig:siam.double.occupation}a and
\ref{fig:siam.double.occupation}b, we show MCTDH data for the SIAM with four bath sites at
$U=2$ and $U=6$. In both cases, the MCTDH results for $N_\mathrm{SPF}^\mathrm{b}=16$ (orange
lines) correspond to the full configuration interaction (TDCI) result and thus perfectly lie
on top of the ED curves. Since the dynamics starts from the atomic limit with a singly-occupied
impurity at $t=0$, the double occupation is initially zero and then becomes finite and oscillatory;
note that the density on the impurity site is a constant of motion by construction of the complex
hopping matrix elements $V_{0l}^\sigma(t)$. For $N_\mathrm{SPF}^\mathrm{b}<16$ the MCTDH results
are approximate, and we generally find that convergence towards ED [by increasing the number of
SPFs] is harder to reach as $U$ decreases. This behavior can be attributed to the fact that, during
the time evolution at small $U$, the inter-site hopping of electrons (i.e., the influence of
$H_\mathrm{hyb}$ in Eq.~(\ref{eq:ham.siam}))  is more pronounced. Consequently, the wave function
expands to a larger area in configuration space which requires an increased number of time-adjusted
SPFs $|\varphi_j(t)\rangle$ to optimally cover the support of $|\Psi(t)\rangle$. For strong coupling
(large $U$) on the contrary, the wave function implies relatively weak inter-coordinate correlation
such that convergence can be reached faster.

In summary, we expect that MCTDH can accurately capture the time evolution of the nonequilibrium
impurity model in the moderate to strong coupling regime, where $U$ is larger than the kinetic energy.
Moreover, it is important to note that the partition of spin-orbitals into combined physical modes
can affect the performance of MCTDH~\cite{vendrell:11}. When a combined mode contains both 
initially empty and initially unoccupied bath orbitals, the initial phase of the dynamics already
involves a larger number of electronic configurations. Thus one may need a higher-dimensional basis
to achieve observables of similar quality. A more favorable partition scheme is to group spin-orbitals
with the criterion that all bath spin-orbitals of a combined mode are initially either empty or fully
occupied. This guarantees that only a small set of possible electronic configurations can be accessed
within the projected Fock space of a certain combined mode.

\begin{figure}[t]
 \includegraphics[width=0.485\textwidth]{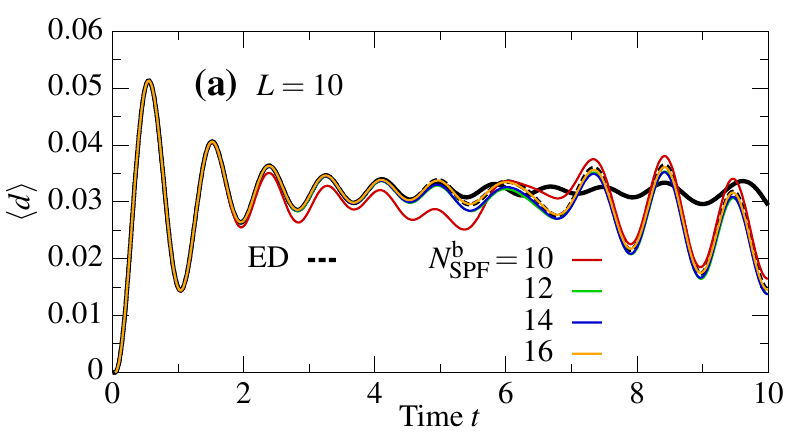}
 \includegraphics[width=0.485\textwidth]{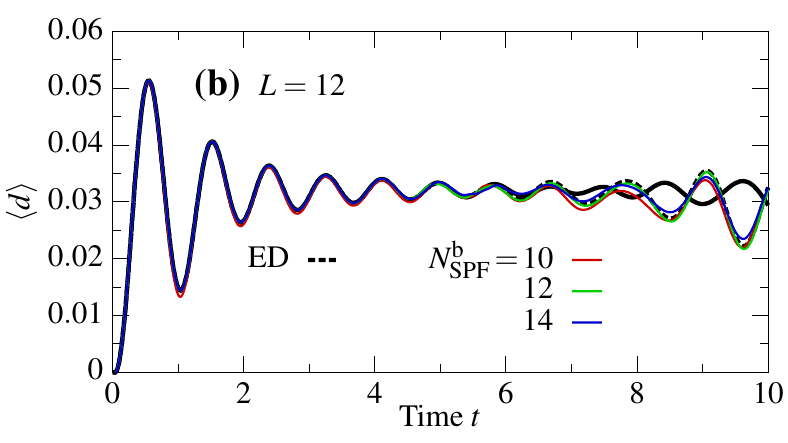}
 \includegraphics[width=0.485\textwidth]{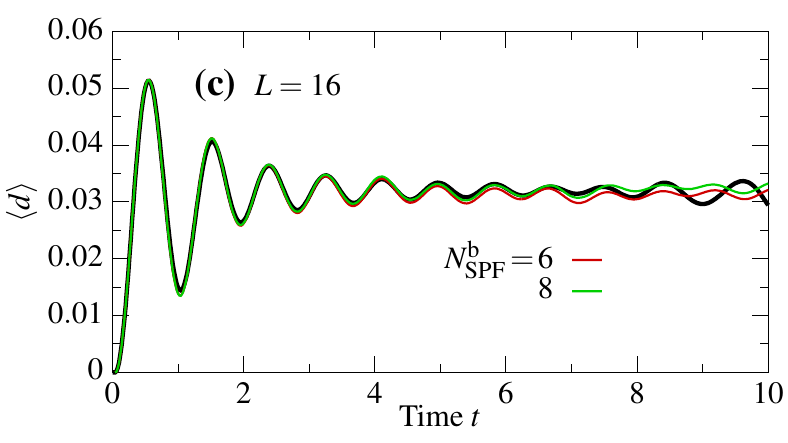}
 \caption{(Color online) Comparison of MCTDH-SQR with four combined modes ($N_\mathrm{CM}=4$) and
 $N_\mathrm{SPF}$ single-particle functions to exact diagonalization (ED) at an on-site interaction
 of $U=10$. Shown is the time-dependent impurity double occupancy $\langle d\rangle(t)$ for the SIAM with 
 \textbf{(a)}~$L=10$, \textbf{(b)}~$L=12$ and \textbf{(c)}~$L=16$ bath orbitals. The black solid
 line, showing the ED result for $L=14$, acts as a reference to determine the maximum time 
 $t_\mathrm{max}$ in Fig.~\ref{fig:scaling}.}
 \label{fig:siam.double.occupation.scaling}
\end{figure}

\subsubsection{Increase of configuration space with time}
\label{subsec:typ.siam.scaling}

We now attempt to estimate the size of the configuration space needed to access a certain maximum
time. For this analysis we restrict ourselves to the case of strong coupling where MCTDH converges
most rapidly ($U=10$). The configuration space is determined by two contributions: (i) the number
of bath orbitals  $L(t_\text{max})$ needed to accurately represent the dynamics up to $t=t_\mathrm{max}$
(cf.~Sec.~\ref{sec:dmft}), and (ii) a possible reduction of the configuration space with respect to
$D_\mathrm{H}(L(t_\mathrm{max}))$ by MCTDH-SQR. 
 
\begin{figure}[t]
 \includegraphics[width=0.485\textwidth]{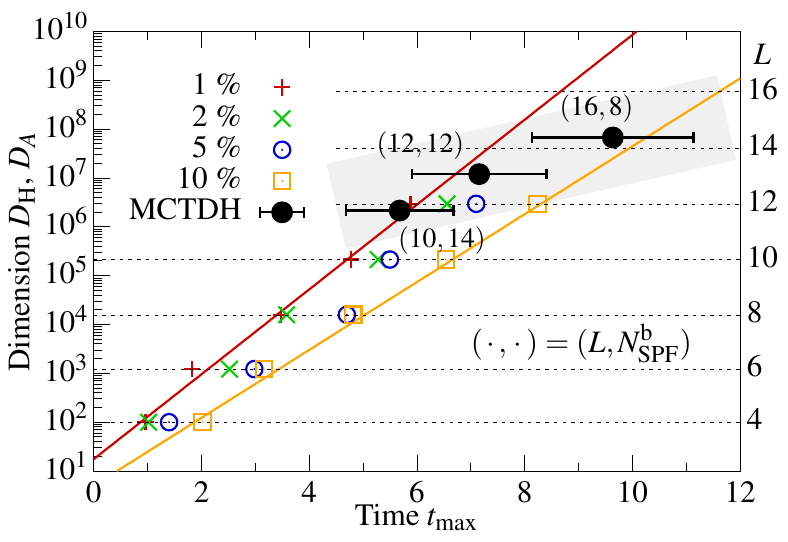}
 \caption{(Color online) Maximum physical time $t_\mathrm{max}$ that can be reached in the
 time evolution of the SIAM with exact diagonalization (colored symbols) and MCTDH-SQR (black
 dots), plotted against the required Hilbert space dimension $D_\mathrm{H}(L)$ for exact diagonalization,
 or the size $D_A(L,N_\mathrm{SPF})$ of the MCTDH tensor. The accessible time $t_\text{max}$
 for a given configuration ($L$,$N_\mathrm{SPF}$) is defined by allowing for a maximum error between $1$ and
 $10$ percent in the double occupation $\langle d\rangle(t_\mathrm{max})$; compare with Figs.~\ref{fig:siam.double.occupation.scaling}a-c.
 The error bars for the MCTDH results are taken from the (extrapolated) difference of the ED data for $1$\%
 and $10$\% deviation, cf.~the red and orange lines.}
 \label{fig:scaling}
\end{figure}

We first determine the configuration space needed within the ED approach. In all panels of
Fig~\ref{fig:siam.double.occupation.scaling}, the black solid line indicates the dynamics for the
bath which is approximated by $14$ sites, which is the largest system size accessible with ED in
our implementation. Comparing these reference data with ED results for smaller $L$ (e.g., the black
dashed lines in Figs.~\ref{fig:siam.double.occupation.scaling}a and \ref{fig:siam.double.occupation.scaling}b
for $L=10$ and $L=12$, respectively), we can extract a maximum physical  time $t_\mathrm{max}(L)$
which can be reached in the calculation with a certain computational effort, measured by the
corresponding Hilbert space dimension $D_\mathrm{H}(L(t_\mathrm{max}))$. The colored symbols
in Fig.~\ref{fig:scaling} indicate exponential scaling between $t_\mathrm{max}$ and $D_\mathrm{H}$
for exact diagonalization, where $t_\mathrm{max}$ is determined by allowing for
a maximum deviation of $1$\% (red crosses) and $10$\% (orange squares) from the $L=14$ reference data.

From the plot it becomes clear that it is exponentially hard to reach long times with a Hamiltonian-based
representation of a DMFT bath. Therefore it is an interesting question whether a MCTDH partition scheme with fewer
and optimally time-evolving SPFs can lead to a more favorable scaling behavior. We obtain indications for
this by analyzing the MCTDH results of Fig.~\ref{fig:siam.double.occupation.scaling} for a minimum number of
SPFs for which $\langle d\rangle(t)$ is still satisfactorily described within an error of about $5$\%. While
in panels Figs.~\ref{fig:siam.double.occupation.scaling}a and \ref{fig:siam.double.occupation.scaling}b we can
directly compare to the corresponding ED result with the same number of bath orbitals (see the black dashed
lines), in Fig.~\ref{fig:siam.double.occupation.scaling}c we only have the $L=14$ data as reference; here we
estimate a maximum time of about $t_\mathrm{max}=9$ up to which the oscillation of the double occupancy for
$N_\mathrm{SPF}^\mathrm{b}=8$ is still decaying as function of time.

The result of the analysis is presented by the black dots labeled by $(L,N_\mathrm{SPF}^\mathrm{b})$
in Fig.~\ref{fig:scaling}. Indeed, we find a deviating scaling for MCTDH which roughly follows the 
delineated gray band as function of $t_\mathrm{max}$. With eight SPFs, the calculation for $L=16$ also 
marks the first point where the size of the MCTDH $A$-vector ($D_A\approx6.7\times10^{7}$)
is smaller than the size of corresponding Hilbert space dealt with in the exact diagonalization
($D_\mathrm{H}\approx5.9\times10^8$).

\subsection{Impurity Green's function}
\label{subsec:imp.gfct}
For a successful implementation as out-of-equilibrium impurity solver, MCTDH must be capable to access
the two-time Green's function $G_{0\sigma}(t,t')$ on the impurity site of the SIAM, from which the hybridization
function is determined in a self-consistent manner. From this local Green's function one can then also obtain,
e.g., the self-energy of the system, the time-dependent momentum distribution or spectroscopic observables of
pump-probe experiments~\cite{aoki:14}.

\begin{figure}[t]
 \includegraphics[width=0.485\textwidth]{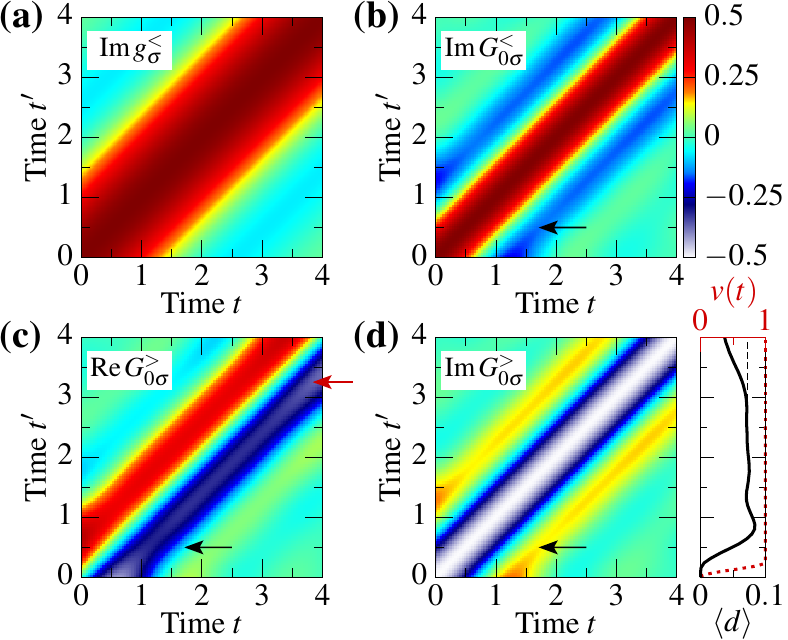}
 \caption{(Color online) \textbf{(a)}~Imaginary part of the Green's function $g^<_\sigma(t,t')$ of Eq.~(\ref{eq:bathgf})
 which is used to compute the initial guess for the hybridization function in the first DMFT iteration. Panels~\textbf{(b)}-\textbf{(d)}: 
 Self-consistent results for the local impurity Green's function $G_{0\sigma}(t,t')$ as obtained from an MCTDH-SQR calculation
 with $L=8$ bath orbitals in the single-impurity Anderson model; the on-site Coulomb repulsion is $U=4$ 
(note that $G^>_{0\sigma}=-G_{0\sigma}^<$ because of particle-hole symmetry).
 The black arrows indicate the early time domain where the transient dynamics due to the switch-on of the hopping is most
 pronounced. The red arrow in panel \textbf{(c)} points to the formation of small artifacts in the final ``steady'' state which
 are due to the representation of the DMFT bath with finitely many bath orbitals. Furthermore, in panel \textbf{(d)}, the black solid
 line shows the time evolution of the double occupation $\langle d\rangle$ in the system, and the red dotted line indicates
 the switch-on of the hopping.}
 \label{fig:gfct}
\end{figure}

To demonstrate the general procedure and its feasibility within MCTDH-SQR we follow Ref.~\cite{gramsch:13}
and consider the real-time dynamics of the Hubbard model on the Bethe lattice, starting from the
atomic limit and from a zero-temperature initial state ($T=0$). More precisely, we fix the on-site
interaction to $U=4$ and study the dynamics of the paramagnetic phase at half-filling when the
nearest-neighbor hopping in the infinite-dimensional lattice is ramped up from zero to $v(t_1)=1$
with a cosine-shaped profile (see the red dotted line in Fig.~\ref{fig:gfct}d); in the Hubbard Hamiltonian~(\ref{eq:ham})
we thus consider $t_{ij}(t)=\delta_{\langle ij\rangle}v(t)/\sqrt{Z}$ in the limit of infinite
coordination number $Z$.

The DMFT action of the lattice Hubbard model is mapped onto a SIAM with an initial state as described
in Sec.~\ref{subsec:typ.siam-model}, i.e., it contains an equal number of empty and doubly-occupied 
bath sites with energy $\epsilon_l=0$ and a singly-occupied impurity. The hopping parameters 
$V_{0l}^\sigma(t)$ are spin-independent and are determined self-consistently via the bath hybridization
function $\Lambda_{\sigma}(t,t')=v(t')G_{0\sigma}(t,t')v(t')$, where $G_{0\uparrow}=G_{0\downarrow}$
for all times on the contour. To generate an initial guess for $\Lambda_\sigma(t,t')$ we use the 
Green's function of Eq.~(\ref{eq:bathgf}), compare with Fig.~\ref{fig:gfct}a.

Given the time-dependent MCTDH wave function $\ket{\Psi(t)}$ of the SIAM for $N=N_\uparrow+N_\downarrow$
particles, the two independent (lesser and greater) components of the impurity Green's function, $G_{0\sigma}^>$
and $G_{0\sigma}^<$, can be computed as the overlaps
\begin{align}
 G_{0\sigma}^>(t,t')&=-\mathrm{i}\langle\Psi(t)|\Phi^>(t,t')\rangle\,,&\\
 G_{0\sigma}^<(t,t')&=\mathrm{i}\langle\Psi(t')|\Phi^<(t',t)\rangle\,,\nonumber
\end{align}
where the states $\ket{\Phi^\gtrless(t,t')}$ are defined by $\ket{\Phi^>(t,t')}=c_{0\sigma}U(t,t')c^{\dagger}_{0\sigma}\ket{\Psi(t')}$, 
$\ket{\Phi^<(t,t')}=c_{0\sigma}^{\dagger}U(t,t')c^{}_{0\sigma}\ket{\Psi(t')}$ and $U(t,t')=T_\mathrm{t}\mathrm{e}^{-i\int_{t'}^{t}\mathrm{d}s\,H'(s)}$ 
denotes the time-evolution operator for the impurity model (\ref{eq:ham.siam}). In practice, we evaluate
the two-time Green's functions as
\begin{eqnarray}
 G_{0\sigma}^>(t,t')&=-\mathrm{i}\langle\Xi^{>}(t)|\Xi^>(t')\rangle\,,&\\
 G_{0\sigma}^<(t,t')&=\mathrm{i}\langle\Xi^<(t')|\Xi^<(t)\rangle\,,\nonumber
\end{eqnarray}
where $\ket{\Xi^>(t)}=U(0,t)\cds{0\sigma}\ket{\Psi(t)}$ and $\ket{\Xi^<(t)}=U(0,t)\ccs{0\sigma}\ket{\Psi(t)}$
are the associated $(N+1)$- and $(N-1)$-particle wave functions.

As the half-filled Hubbard model we start from is particle-hole symmetric, but the SIAM with
spin imbalanced occupation is not, we use an adapted initial state which is a superposition of
two degenerate states: One has a spin-up electron occupying the impurity site, and the other has
a spin-down electron on the impurity site. An alternative scheme which we have also implemented
to restore particle-hole symmetry is to first construct Green's functions $G^A(t,t')$ and $G^B(t,t')$
with interchanged particle numbers (i.e., $N_{\uparrow}\leftrightarrow N_{\downarrow}$), and
then to average over the two Green's functions according to $G(t,t')=\tfrac{1}{2}[G^A(t,t')+G^B(t,t')]$.

In Figs.~\ref{fig:gfct}b-d we present results for the self-consistent impurity Green's function where the 
hybridization function has been approximated on a time window $[0,4]$ by a SIAM with $L=8$ bath orbitals.
We clearly see that while the density in the system, $\langle n_\sigma \rangle(t)=\mathrm{Im}\,G^<_\sigma(t,t)=0.5$,
is a constant of motion, the time-off-diagonal components of the Green's function containing the spectral
information develop as function of the two times (see the black arrows). Moreover, for times $t,t'\gtrsim1.5$
where the double occupation in the system approaches a stationary value, also the Green's functions attain
quasi static structure as function of the physical (center of mass) time $(t+t')/2$.

To bring the results to convergence, we have implemented the self-consistency loop in two ways, either
iterating on the full $(t,t')$-mesh or using the time propagation scheme described in Ref.~\cite{gramsch:13}.
While the former approach was simpler to implement, the latter is found to be much more efficient because
the self-consistency is established for each time slice separately allowing for essentially fewer iterations.
Finally, we remark that the tiny changes in the Green's function at later times ($t\gtrsim3$) are due to the
discretization of the DMFT bath with only eight bath sites, see, e.g., the red arrow in Fig.~\ref{fig:gfct}c
and compare to the time evolution of the double occupancy in Fig.~\ref{fig:gfct}d which also deviates from the
steady state (dashed line) for times $t>3$.

%%%%%%%%%%%%%%%%%%%%%%%%%%%%%%%%%%%%%%%%%%%%%%%%%%%%%%%
% Conclusion
%%%%%%%%%%%%%%%%%%%%%%%%%%%%%%%%%%%%%%%%%%%%%%%%%%%%%%%
\section{Conclusion}
\label{sec:conclusion}
In this work we have implemented and benchmarked a solution of the impurity problem 
of nonequilibrium dynamical mean-field theory (DMFT) based on the multiconfiguration
time-dependent Hartree (MCTDH) method. The MCTDH method provides a variationally optimized
representation of a time-dependent (fermionic or bosonic) wave function, which can reduce
the dimension of the underlying basis function space by several orders of magnitude. The
resulting compression of the wave function is a crucial feature to overcome the notorious
exponential scaling barrier in the Hamiltonian representation of the DMFT action that
hinders the access of long simulation times.

For the time-dependent single-impurity Anderson model (SIAM), which represents
the core component of the Hamiltonian-based DMFT approach out of equilibrium,
we have been able to show that MCTDH can indeed go beyond the capability of
exact diagonalization for sufficiently strong Coulomb interactions. 
For SIAMs with a small number of bath sites, an exact solution is more favorable
than MCTDH, because the latter is implemented in second quantization representation 
(MCTDH-SQR) where the state vector is defined in Fock space and contains redundant 
electronic configurations which are unphysical for simulating a system with a given 
number of electrons (and spin-orbitals). For large systems, as needed to solve the 
DMFT problem at long times, MCTDH can become favorable. Using the concept of mode 
combination, we have provided the dynamics of a SIAM, which describes a typical DMFT 
bath with $L=16$ bath orbitals and is not feasible to be solved by exact diagonalization 
at reasonable computational cost. This calculation marks the onset of a regime in which
the state vector of MCTDH scales more favorably than the Hilbert space with the maximum 
physical time that can be accessed in the simulation.

Moreover, we have illustrated the feasibility of the MCTDH-SQR algorithm to yield
the time-dependent observables as well as the self-consistent real-time Green's functions
for a generic Hubbard-type lattice problem, which shows the potential of the method to act as full 
DMFT impurity solver. Although the efficiency to directly compute the Green's function 
with the standard Heidelberg MCTDH package for system sizes as large as $L\approx16$ must still be
proven, there are no conceptual difficulties to upscale the scope of the method.

In order to push the applicable regime of the MCTDH-based impurity solver to even
larger SIAMs and hence to even longer time scales, it is very promising to 
extend the approach to multi-layer MCTDH~\cite{wang:09,manthe:08,vendrell:11},
by which the scaling barrier is expected to be more efficiently overcome.
Thus, to find an optimal tree-tensor network decomposition of the fermionic SIAM
wave function within the ML-MCTDH scheme is the main path for future work.

%%%%%%%%%%%%%%%%%%%%%%%%%%%%%%%%%%%%%%%%%%%%%%%%%%%%%%%
% Acknowledgments
%%%%%%%%%%%%%%%%%%%%%%%%%%%%%%%%%%%%%%%%%%%%%%%%%%%%%%%
\acknowledgments
Z.L.~and O.V.~acknowledge financial support from the Hamburg Centre for Ultrafast
Imaging (CUI). The authors thank Michael Bonitz, Stephen Clark, Rainer H\"artle,
Marcus Kollar, and Hans-Dieter Meyer for intensive discussions.

%%%%%%%%%%%%%%%%%%%%%%%%%%%%%%%%%%%%%%%%%%%%%%%%%%%%%%%
% Conclusion
%%%%%%%%%%%%%%%%%%%%%%%%%%%%%%%%%%%%%%%%%%%%%%%%%%%%%%%

%%%%%%%%%%%%%%%%%%%%%%%%%%%%%%%%%%%%%%%%%%%%%%%%%%%%%%%%
% DOCUMENT END 
%%%%%%%%%%%%%%%%%%%%%%%%%%%%%%%%%%%%%%%%%%%%%%%%%%%%%%%%

\begin{thebibliography}{100}
\expandafter\ifx\csname natexlab\endcsname\relax\def\natexlab#1{#1}\fi
\expandafter\ifx\csname bibnamefont\endcsname\relax
  \def\bibnamefont#1{#1}\fi
\expandafter\ifx\csname bibfnamefont\endcsname\relax
  \def\bibfnamefont#1{#1}\fi
\expandafter\ifx\csname citenamefont\endcsname\relax
  \def\citenamefont#1{#1}\fi
\expandafter\ifx\csname url\endcsname\relax
  \def\url#1{\texttt{#1}}\fi
\expandafter\ifx\csname urlprefix\endcsname\relax\def\urlprefix{URL }\fi
\providecommand{\bibinfo}[2]{#2}
\providecommand{\eprint}[2][]{\url{#2}}

\bibitem{cavalieri:07}
A.L.~Cavalieri, N.~M\"{u}ller, T.~Uphues, V.S.~Yakovlev, A.~Baltu$\check{\rm s}$ka, B.~Horvath, B.~Schmidt, L.~Bl\"{u}mel, R.~Holzwarth, S.~Hendel, 
M.~Drescher, U.~Kleineberg, P.M.~Echenique, R.~Kienberger, F.~Krausz, and U.~Heinzmann, Nature \textbf{449}, 1029 (2007).

\bibitem{wall:11}
S.~Wall, D.~Brida, S.R.~Clark, H.P.~Ehrke, D.~Jaksch, A.~Ardavan, S.~Bonora, H.~Uemura, Y.~Takahashi, T.~Hasegawa, H.~Okamoto, G.~Cerullo and A.~Cavalleri, Nature Phys.~\textbf{7}, 114 (2011).

\bibitem{perfetti:06}
L.~Perfetti, P.A.~Loukakos, M.~Lisowski, U.~Bovensiepen, H.~Berger, S.~Biermann, P.~S.~Cornaglia, A.~Georges, and M.~Wolf, Phys.~Rev.~Lett.~\textbf{97}, 067402 (2006).

\bibitem{iwai:03}
S.~Iwai, M.~Ono, A.~Maeda, H.~Matsuzaki, H.~Kishida, H.~Okamoto, and Y.~Tokura, Phys.~Rev.~Lett.~\textbf{91}, 057401 (2003).

\bibitem{schmitt:08}
F.~Schmitt, P.S.~Kirchmann, U.~Bovensiepen, R.G.~Moore, L.~Rettig, M.~Krenz, J.-H.~Chu, N.~Ru, L.~Perfetti, D.H.~Lu, M. Wolf, I.R.~Fisher, and Z.-X.~Shen, Science \textbf{321}, 1649 (2008).

\bibitem{schollwoeck:05}
U.~Schollw\"ock, Rev.~Mod.~Phys.~\textbf{77}, 259 (2005).

\bibitem{schmidt:02}
P.~Schmidt and H.~Monien, arXiv:cond-mat/0202046 (2002).

\bibitem{freericks:06}
J.K.~Freericks, V.M.~Turkowski, and V.~Zlati\'{c}, Phys.~Rev.~Lett.~\textbf{97}, 266408 (2006).

\bibitem{aoki:14}
H.~Aoki, N.~Tsuji, M.~Eckstein, M.~Kollar, T.~Oka, and P.~Werner, Rev.~Mod.~Phys.~\textbf{86}, 779 (2014).

\bibitem{georges:96}
A.~Georges, G.~Kotliar, W.~Krauth, and M.J. Rozenberg, Rev.~Mod.~Phys.~\textbf{68}, 13 (1996).

\bibitem{eckstein:09}
M.~Eckstein, M.~Kollar, and P.~Werner, Phys.~Rev.~Lett.~\textbf{103}, 056403 (2009).

\bibitem{eckstein:10.nca}
M.~Eckstein and P.~Werner, Phys.~Rev.~B \textbf{82}, 115115 (2010).

\bibitem{tsuji:14}
N.~Tsuji and P.~Werner, Phys.~Rev.~B \textbf{88}, 165115 (2013).

\bibitem{gramsch:13}
C.~Gramsch, K.~Balzer, M.~Eckstein, and M.~Kollar, Phys.~Rev.~B \textbf{88}, 235106 (2013).

\bibitem{szaboostlund96}
A.~Szabo and N.A.~Ostlund, \textit{Modern Quantum Chemistry: Introduction to Advanced Electronic Structure Theory} (Dover Publications Inc., Mineola, NY, 1996).

\bibitem{balzer:14}
K.~Balzer and M.~Eckstein, Phys.~Rev.~B \textbf{89}, 035148 (2014).

\bibitem{hofmann:13}
F.~Hofmann, M.~Eckstein, E.~Arrigoni, and M.~Potthoff, Phys.~Rev.~B \textbf{88}, 165124 (2013).

\bibitem{arrigoni:13}
E.~Arrigoni, M.~Knap, and W.~von~der Linden, Phys.~Rev.~Lett.~\textbf{110}, 086403 (2013).

\bibitem{verstraete:04}
F.~Verstraete and J.I.~Cirac, arXiv:cond-mat/0407066v1 (unpublished).

\bibitem{vidal:07}
G.~Vidal, Phys.~Rev.~Lett.~\textbf{99}, 220405 (2007).

\bibitem{murg:07}
V.~Murg, F.~Verstraete, and J.I.~Cirac, Phys.~Rev.~A \textbf{75}, 033605 (2007).

\bibitem{beck:00}
M.H.~Beck, A.~J{\"a}ckle, G.A. Worth, and H.-D.~Meyer, Phys.~Rep.~\textbf{324}, 1 (2000).

\bibitem{meyer:09}
H.-D.~Meyer, F.~Gatti, and G.~Worth (eds.), \textit{Multidimensional Quantum Dynamics: MCTDH theory and its applications} (Wiley-VCH, 2009).

\bibitem{wang:09}
H.~Wang and M.~Thoss, J.~Chem.~Phys.~\textbf{131}, 024114 (2009).

\bibitem{manthe:08}
U.~Manthe, J.~Chem.~Phys.~\textbf{128}, 164116 (2008).

\bibitem{vendrell:11}
O.~Vendrell and H.-D.~Meyer , J.~Chem.~Phys.~\textbf{134}, 044135 (2011).

\bibitem{albrecht:12}
K.F.~Albrecht, H.~Wang, L.~M\"uhlbacher, M.~Thoss, and A.~Komnik, Phys.~Rev.~B \textbf{86}, 081412 (2012).

\bibitem{wang:13.jchemphys}
H.~Wang and M.~Thoss, J.~Chem.~Phys.~\textbf{138}, 134704 (2013).

\bibitem{kotliar:04}
G.~Kotliar and D.~Vollhardt, Physics Today \textbf{57}, 53 (2004).

\bibitem{lichtenstein:98}
A.I.~Lichtenstein and M.I.~Katsnelson, Phys.~Rev.~B \textbf{57}, 6884  (1998).

\bibitem{anisimov:97}
V.I.~Anisimov, A.I.~Poteryaev, M.A.~Korotin, A.O.~Anokhin, and G.~Kotliar, Journal of Physics: Condensed Matter \textbf{9}, 7359 (1997).

\bibitem{imada:98}
M.~Imada, A.~Fujimori, and Y.~Tokura, Rev.~Mod.~Phys.~\textbf{70}, 1039 (1998).

\bibitem{keldysh:64}
L.V.~Keldysh, Zh.~Eksp.~Teor.~Fiz.~\textbf{47}, 1515 (1964) [Sov.~Phys.~JETP \textbf{20},1018 (1965)].

\bibitem{eckstein:10.epjst}
M.~Eckstein, A.~Hackl, S.~Kehrein, M.~Kollar, M.~Moeckel, P.~Werner, and F.A.~Wolf, Eur.~Phys.~J.~Special Topics \textbf{180}, 217 (2010).

\bibitem{mctdh:package}
G.A. Worth, M.H. Beck, A.~J{\"a}ckle, and H.-D. Meyer, The {MCTDH} {P}ackage, {V}ersion 8.2, (2000). H.-D.~Meyer, {V}ersion 8.3 (2002), {V}ersion 8.4 (2007), {V}ersion 8.5 (2014). {S}ee http://mctdh.uni-hd.de.

\bibitem{meyer:90}
H.-D.~Meyer, U.~Manthe, and L.S.~Cederbaum, Chem.~Phys.~Lett.~\textbf{165}, 73 (1990).

\bibitem{tucker:66}
L.~Tucker, Psychometrika \textbf{31}, 279 (1966).

\bibitem{kolda:09}
T.G.~Kolda and B.W.~Bader, SIAM Review \textbf{51}, 455 (2009).

\bibitem{wang:08.njp}
H.~Wang and M.~Thoss, New J.~Phys.~\textbf{10}, 115005 (2008).

\bibitem{caillat:05}
J.~Caillat, J.~Zanghellini, M.~Kitzler, O.~Koch, W.~Kreuzer, and A.~Scrinzi, Phys.~Rev.~A \textbf{71}, 012712 (2005).

\bibitem{nest:09}
M.~Nest, Chem.~Phys.~Lett.~\textbf{472}, 171 (2009).

\bibitem{ulusoy:12}
I.S.~Ulusoy and M.~Nest, J.~Chem.~Phys.~\textbf{136}, 054112 (2012).

\bibitem{zhang:14}
J.M.~Zhang and M.~Kollar, Phys.~Rev.~A \textbf{89}, 012504 (2014).

\bibitem{hochstuhl:14}
D.~Hochstuhl, C.M.~Hinz, and M.~Bonitz, Eur.~Phys.~J.~Special Topics \textbf{223}, 177 (2014).

\bibitem{alon:07}
O.E.~Alon, A.I.~Streltsov, and L.S.~Cederbaum, J.~Chem.~Phys.~\textbf{127}, 154103 (2007).

\bibitem{alon:08}
O.~Alon, A.~Streltsov, and L.~Cederbaum, Phys.~Rev.~A \textbf{77}, 033613 (2008).

\bibitem{cao:13}
L.~Cao, S.~Kr\"onke, O.~Vendrell, and P.~Schmelcher, J.~Chem.~Phys.~\textbf{139}, 134103 (2013).

\end{thebibliography}
\end{document}